\documentstyle[twocolumn,prb,aps,graphicx]{revtex}

\def\AmS{{\protect\the\textfont2
        A\kern-.1667em\lower.5ex\hbox{M}\kern-.125emS}}

\makeatletter
\def\@pnumwidth{2em}
\makeatother
\begin{document}
\DeclareGraphicsExtensions{.eps,.jpg,.pdf,.mps,.png} \draft
\twocolumn[\hsize\textwidth\columnwidth\hsize\csname
@twocolumnfalse\endcsname
\title{A modified dual-slope method for heat capacity measurements of
condensable gases}
\author{S.\ Pilla, J.\ A.\ Hamida, and N.\ S.\ Sullivan}
\address{Department of Physics, University of Florida,
 Gainesville, FL 32611}
\date{\today}
\maketitle

\begin{abstract}
A high resolution non-adiabatic method for measuring the heat
capacity ($C_v$) of bulk samples of condensable gases in the
range 7.5-70 K is described. In this method $C_v$ is evaluated by
directly comparing the heating and cooling rates of the sample
temperature for two algebraically independent heat pulse
sequences without explicit use of the thermal conductance between
sample and thermal bath. A fully automated calorimeter for rapid
measurement of $C_v$ of molecular solids utilizing this technique
is presented.
\end{abstract}

\pacs{}
]
\narrowtext
\section{INTRODUCTION}
The adiabatic heat-pulse calorimetry has been widely used to
investigate the thermodynamic properties of materials for more
than a century. The applied principle, $C=\triangle Q/\triangle
T$, which describes the heat capacity $C$ of the sample, is
determined by the pulse heat $\triangle Q$ supplied to the sample
under adiabatic conditions and the temperature rise $\triangle T$,
is well known.\cite{Eucken,Nernst} Due to the inherent simplicity
as well as the general applicability independent of the sample
thermal conductivity, this method is the most favored choice for
heat capacity measurements of condensable gases which have poor
thermal conductivity in their low temperature solid
phase.\cite{Alkhafaji,Bagatskii,Ward} Although the traditional
adiabatic calorimetry has high precision and can be used to
determine the latent heat at strong first-order transitions, it
is very difficult to achieve the resolution needed to characterize
the temperature dependence of $C_v(T)$ (or $C_p(T)$) close to the
critical temperature $T_c$ for a second-order transition. Also,
because of the inherent limitations on achieving the ideal
adiabatic conditions at low temperatures and the long time
required to cover a few tens of kelvin temperature range with
reasonable number of data points, new user friendly non-adiabatic
techniques
\cite{Sullivan,Bachmann,Riegel,Xu,Pecharsky,Yao,Hwang,Jin} with
excellent sensitivity are needed to study the heat capacity of
condensable gases. Among these new techniques, the most sensitive
method is the ac method devised by Sullivan and
Seidel.\cite{Sullivan} While the ac method allows one to obtain
accurate values of $C_v$, it suffers from the fact that it
normally must be used for small samples at low ac frequencies and
over limited temperature ranges (typically $T < 20$ K). In order
to keep the sample in equilibrium with the heater and thermometer
during one cycle, the time constant for thermal relaxation
($\tau_1$) of the sample plus sample holder (or calorimeter) to
the thermal reservoir must be short compared to the period of the
driving flux. In addition, the samples's internal equilibrium
time constant ($\tau_2$) with the sample holder (or calorimeter)
must be short compared to $\tau_1$. It is this latter constraint
which restricts the applicability of this method to small samples
with sufficiently large thermal conductivity such as pure metals
below 20 K.

A slightly less sensitive method involving a fixed heat input
followed by a temperature decay measurement is the relaxation
method. \cite{Bachmann,Forgan,Schwall} In this method the sample
is raised to an equilibrium temperature above the thermal
reservoir and then allowed to relax to the reservoir temperature
without heat input. A recent improvement of this technique
utilizing advanced numerical methods was provided by Hwang $et$
$al.$\cite{Hwang} The relaxation method requires extensive
calibration of the heat losses of the sample as a function of
temperature difference between the reservoir and the final sample
temperature and relies on accurate, smooth temperature
calibration of thermometers. The technique depends heavily on the
numerical techniques used to determine many equilibrium heat
losses during the heating portion of the relaxation cycle to
obtain good temperature resolution of the heat capacity changes.
Though Hwang $et$ $al.$\cite{Hwang} addressed the problems
arising from large $\tau_2$, this method still fails to provide
accurate $C_v$ values above $\sim$20 K for large samples with
poor thermal conductivity.

A variation of the relaxation method called dual-slope method,
was first discussed by Riegel and Weber.\cite{Riegel} In this
method an extremely weak thermal link to the reservoir is used and
the temperature of the sample is recorded over a 10 h.\ cycle
while heating at constant power for one half of the cycle, and
then allowing the sample relax with zero heat input. The heat loss
to the reservoir and surroundings can be eliminated from the
calculation of $C_v(T)$ using this technique provided the sample,
sample holder, and the thermometer are always in equilibrium with
each other (the reason for long 10 h.\ cycle to cover 3 K range),
and the reservoir temperature is held constant over the 10 h.\
cycle. A further modification of this technique is a hybrid
between the ac method and the dual-slope method,\cite{Xu} which
reduces the duration of the cycle to about two hours. Although
the dual-slope method is very elegant and easy to implement, the
technique has only been implemented for small samples (typically
less than 0.5 g) with good thermal conductivity and at
temperatures less than $\sim$20 K. The success of this method
depends heavily on achieving very good thermal equilibrium
between the sample, sample holder, and the thermometer, and for
large samples with poor thermal conductivity (i.e., large
$\tau_2$) this method may also fail when we consider only the
first order approximation of the heat balance equations of Riegel
and Weber.\cite{Riegel,Xu,Hwang}

In the case of condensable gases, the sample size must be at
least a few grams to reduce the spurious effect of sample
condensed in the fill line and the heat leak through this tube.
Due to the low thermal conductivity of these samples, in
particular for powdered samples which may not wet the calorimeter
walls, the $\tau_2$ can be very large leading to the failure of
most of the above techniques. The present article describes a
modified dual-slope method in which $C_v$ is evaluated by
directly comparing the heating and cooling rates of the sample
temperature for two algebraically independent heat pulse
sequences without explicit use of the thermal conductance between
sample and thermal bath. For the specific geometry of the
calorimeter that we used, which is most suitable for the heat
capacity measurements of condensable gases in the presence of
external electric or magnetic fields, higher order heat balance
equations are calculated, which can easily be adapted for other
sample configurations as well. Because of the explicit
consideration of the higher order equations, the problem of large
$\tau_2$ is naturally addressed and during the heating and
cooling cycles one does not require the sample to be in
equilibrium with the thermometer, or in other words, the
thermometer need not necessarily record the actual sample
temperature. Due to this freedom we can obtain $C_v$ of samples
with poor thermal conductivity very rapidly.

\section{Theory}
\begin{figure}
\begin{center}
\leavevmode \includegraphics[width=.75\linewidth]{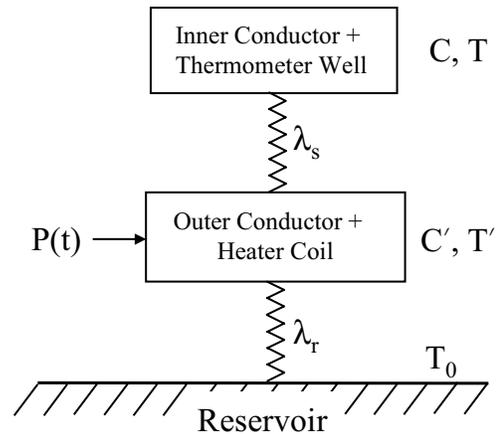}
\end{center}
\caption{Simplified diagram of the thermal system of sample,
sample holder, and thermal bath. The thermal link between the
outer conductor plus heater of heat capacity $C^{\prime}$ at
temperature $T^{\prime}$ and temperature-controlled reservoir
bath at $T_0$ is represented by the thermal conductance
$\lambda_{r}$. $\lambda_{r}$ may be controlled by changing the He
exchange gas pressure. The heater input power $P(t)$ is obtained
from a waveform generator and a suitable power amplifier. The
thermal link (through the sample) between the inner conductor plus
thermometer well of heat capacity $C$ at temperature $T$ and the
outer conductor is represented by the thermal conductance
$\lambda_{s}$.} \label{model}
\end{figure}

To model the thermal response of a heat-pulse calorimeter (Fig.\
\ref{calorimeter}) for heat capacity measurements, with the
schematic diagram shown in Fig. \ref{model}, where the
calorimeter is heated with the application of heater power $P(t)$
to some desired temperature $T_{max}$ and then allowed to
cool-down, we start with the following set of heat balance
equations:
\begin{equation}
P(t)=C^{\prime}\dot{T_h^{\prime}}+\lambda_s(T_h^{\prime}-T_h)+
\lambda_r(T_h^{\prime}-T_0)
\end{equation} and
\begin{equation}
0=C\dot{T_h}+\lambda_s(T_h-T_h^{\prime})-P_0(T_h)
\end{equation} for heating, and
\begin{equation}
0=C^{\prime}\dot{T_c^{\prime}}+\lambda_s(T_c^{\prime}-T_c)+
\lambda_r(T_c^{\prime}-T_0)
\end{equation} and
\begin{equation}
0=C\dot{T_c}+\lambda_s(T_c-T_c^{\prime})-P_0(T_c)
\end{equation} for cooling,
where $C$ and $T$ are the heat capacity and the temperature of
the thermometer well plus the inner conductor respectively,
$C^{\prime}$ and $T^{\prime}$ are those of the outer conductor
plus the heater coil. From these equations we need to obtain the
quantity $(C+C^{\prime})$ which is the net heat capacity of the
calorimeter. Once the sample is condensed in the calorimeter,
both $C$ and $C^{\prime}$ will be modified but the above
equations are still valid. $P_0(T)$ is the parasitic stray heat
due to the conduction of heat through the fill line, the stainless
steel suspension rod, and the thermometer leads as well as
radiation from the pumping line. From the geometry of the
calorimeter (Fig. \ref{calorimeter}) it is reasonable to assume
that this stray heat affects only the inner conductor heat
balance equations (Eqs. 2 \& 4).

Since the thermometer is firmly coupled to the inner conductor
through the thermometer well, the temperature $T$ is what the
thermometer practically records. Hence we need to eliminate
$T^{\prime}$ to obtain $(C+C^{\prime})$. For $T_h=T_c$ Eqs. 1 \&
3 reduce to
\begin{equation}
P(t)=C^{\prime}(\dot{T_h^{\prime}}-\dot{T_c^{\prime}})+(\lambda_s+\lambda_r)
(T_h^{\prime}-T_c^{\prime}).
\end{equation}
From Eqs. 2 \& 4 and their first derivatives we obtain
\begin{equation}
(T_h^{\prime}-T_c^{\prime})=\frac{C}{\lambda_s}(\dot{T_h}-\dot{T_c}),
\end{equation}
\begin{equation}
(\dot{T_h^{\prime}}-\dot{T_c^{\prime}})=
(\dot{T_h}-\dot{T_c})+\frac{C}{\lambda_s}(\ddot{T_h}-\ddot{T_c}).
\end{equation}
After eliminating $T_h^{\prime}$, $T_c^{\prime}$ and their time
derivatives from Eqs. 5, 6, and 7 we obtain
\begin{equation}
P(t)=(C+C^{\prime})(\dot{T_h}-\dot{T_c})+\frac{CC^{\prime}}{\lambda_s}
(\ddot{T_h}-\ddot{T_c})+\frac{C\lambda_r}{\lambda_s}(\dot{T_h}-\dot{T_c}).
\end{equation}

Clearly, Eq.\ (8) reduces to the first order derivations of Riegel
and Weber\cite{Riegel} for $\lambda_s \gg 1$. However, for
condensed gases, in particular for powdered samples, this
condition is never met. In the present geometry, for a
sufficiently weak thermal link $\lambda_r$ where $\lambda_s \gg
\lambda_r$, one can ignore the last term in Eq.\ (8). To solve the
remaining equation exactly for $C+C^{\prime}$, we need to obtain
Eq.\ (8) for two algebraically independent pulse sequences of
$P(t)$.

\section{Experiment}
\begin{figure}
\begin{center}
\leavevmode \includegraphics[width=.6\linewidth]{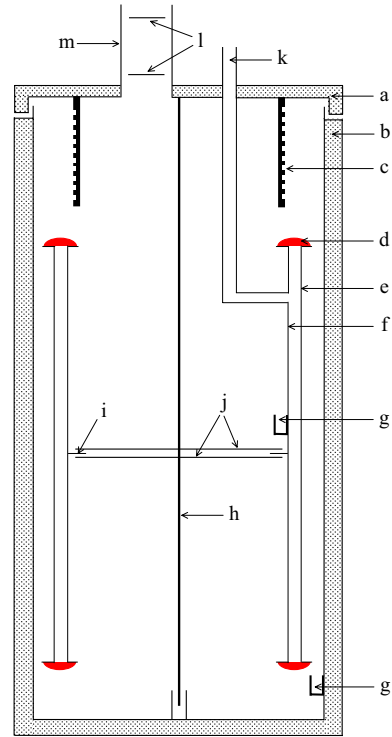}
\end{center}
\caption{Cross section of the thermal reservoir and calorimeter.
(a) Top flange with vacuum feed-throughs, (b) brass vacuum can
with highly uniform heater wire wound on the outer surface, (c)
1.6 cm diameter copper tube with grooves on the outer surface, (d)
STYCAST 2850FT seal,\cite{Pilla} (e) outer conductor with highly
uniform heater wire wound on the outer surface, (f)inner
conductor, (g) copper thermometer wells soldered to the walls,
(h) threaded stainless steel rod, (i) 0.5 mm thick copper ring
soldered to the inner conductor, (j) threaded nylon wings, (k)
sample fill line, (l) copper radiation baffles, (m) stainless
steel tube for support as well as pumping.} \label{calorimeter}
\end{figure}

Figure \ref{calorimeter} shows the design of the calorimeter
optimized for the above technique. The thermal reservoir is made
up of a brass vacuum can (18 cm long, 4 cm diameter) with a
manganin wire wound uniformly on the entire length of the outer
surface. A dilute solution of GE-varnish is used to soak the cloth
insulation of the manganin heater wire to provide good thermal
contact with the vacuum can. A grooved copper tube (4 cm long) is
soldered to the top flange, which can be used for heat sinking the
heater and thermometer wires before connecting them to the
calorimeter. A 122 cm long stainless steel tube (0.96 cm
diameter) is soldered to the top flange to pump the vacuum space
as well as support the vacuum can. copper radiation baffles are
placed inside this tube at regular intervals to reduce the
thermal radiation. In addition the top flange supports up to ten
individual vacuum feed-throughs (not shown), which are thermally
anchored to the flange through STYCAST 2850FT vacuum seal which
has good thermal conductivity at low temperatures. The
calorimeter consists of two concentric, thin walled, gold plated,
OFHC copper cylinders 10.2 cm in length providing a 1 mm
cylindrical shell space for condensing the samples. This shell
space is vacuum sealed at both ends with homemade 1 mm thick
STYCAST 2850FT O-rings.\cite{Pilla} The outer conductor (2.7 cm
inner diameter and 0.2 mm wall thickness) with a brass heater
wire uniformly wound on the entire length of its outer surface
serves as a radial heating element. Uniform winding of the heater
wire is very crucial in eliminating longitudinal heat flow
thereby reducing the temperature gradient along the length of the
calorimeter. For good thermal contact with the outer cylinder, a
dilute solution of GE-varnish was applied to the heater wire. In
the middle of the inner conductor (2.5 cm diameter, 0.4 mm wall
thickness) a thin copper ring as well as a copper thermometer
well are soldered (see Fig. \ref{calorimeter}). The larger wall
thickness (0.4 mm instead of 0.2 mm) for the inner cylinder is
necessary to withstand the pressures generated when the solid
samples melt at high temperatures. The copper ring sandwiched
between two threaded nylon wings supports the calorimeter when
suspended from the top flange with the help of a threaded
stainless steel rod. This arrangement not only reduces the
unwanted heat leaks to the calorimeter but also provides
excellent mechanical stability when the free end of the stainless
steel rod is slipped into the post at the bottom of the vacuum
can (see Fig. \ref{calorimeter}).
The 1.6 mm diameter cupronickel alloy tube serving as the sample
fill line is coiled into a three turn spring (not shown) before
soldering it to the inner cylinder. To be able to apply electric
field across the sample cell when necessary, it is important to
ground the outer conductor (to reduce electrical interference
from the heater current) and apply the high voltage to the inner
conductor. Hence the coiled section of the fill line is isolated
electrically as well as thermally from the remaining length with
the help of a vacuum joint made up of a wider cupronickel tube,
cigarette paper, and STYCAST seal. To assist sample condensation,
the manganin heater wire is wrapped on the fill line tube outside
of the vacuum can. Two factory calibrated germanium resistance
thermometers\cite{SI} placed inside the thermometer wells are
used for recording the temperatures of the calorimeter and the
thermal reservoir. The entire assembly is placed in a second
vacuum can (125 cm long, 5 cm diameter) which can be lowered into
a commercial liquid He dewar. With this arrangement including a
few suitably placed copper radiation baffles at the neck of the
outer vacuum can, a standard 100 liter dewar lasted for almost a
month while heat capacity experiments were carried out in the
$4.2-75$ K range.
\section{RESULTS AND DISCUSSION}
\begin{figure}
\begin{center}
\leavevmode \includegraphics[width=.9\linewidth]{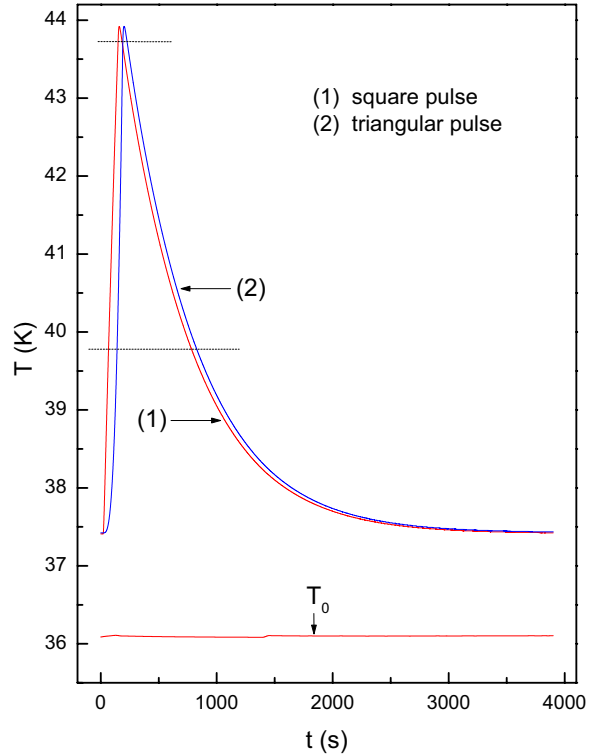}
\end{center}
\caption{Traces of $T(t)$ and $T_0(t)$ for two heat pulses when
the calorimeter is filled with pure N$_2$ sample. Useful interval
for obtaining heat capacity is shown between two dashed lines.
The difference between the trace of $T_0$ and the baseline of $T$
is due to $P_0$.} \label{heatpulse}
\end{figure}

Our observations indicate that due to the high degree of
cylindrical symmetry, even though the calorimeter is rather
unusually long ($\sim$10 cm), thermometers record the expected
temperature values to very high accuracy. Figure \ref{heatpulse}
shows typical temperature excursions when the calorimeter is
filled with pure N$_2$ as a sample. For simplicity we chose square
and triangular voltage pulses for two algebraically independent
$P(t)$ sequences. Both pulses can be easily generated with the
help of a standard signal generator or data acquisition card and
a suitable operational amplifier. Before the heat pulse is
applied, the calorimeter is left to equilibrate with the
surroundings until its temperature trace is horizontal with the
time axis. Typically, when the reservoir temperature is adjusted
to a new value, the calorimeter attained equilibrium in 30 to 60
min. To ensure that the calorimeter always attained equilibrium
from either a higher temperature or a lower temperature for all
pulse sequences (i.e., $T\rightarrow T_0^+$ or $T\rightarrow
T_0^-$), and that small hysteresis of the the thermometer does not
affect the trace, when $T_0$ is set to a higher value, a small
heat pulse is simultaneously applied to the calorimeter so that
its temperature rises above the equilibrium temperature by at
least 2 to 3 K. When such a precaution was not taken, we observed
that the base line of the two traces in Fig.\ \ref{heatpulse}
differed by $\sim$100 mK resulting in spurious results when fit
to Eq.\ (8).

Although we assumed that the last term in Eq.\ (8) will be
negligible for $\lambda_s \gg \lambda_r$, it is clear that for
the long exponential decay section of the traces (below the lower
horizontal line in Fig.\ \ref{heatpulse}) where
$(\dot{T_h}-\dot{T_c})$ can be very large, the above
approximation may not be valid. We observed that for a typical
6.5 K pulse (similar to Fig.\ \ref{heatpulse}), the useful window
of temperature where the above approximation is valid for
obtaining faithful results is only 3.5$\sim$4 K (the section
between the two dashed lines in Fig.\ \ref{heatpulse}). It is
crucial that during the entire pulse sequence, the reservoir
temperature $T_0$ should be constant. With the help of a program
written in LabVIEW,\cite{NI} which controls the voltage across
the reservoir heater through a multifunction AD/DA card\cite{NI}
and with the thermometer output in the negative feedback loop, we
were able to maintain $T_0$ to within 30 mK during the entire
pulse sequence (see Fig.\ \ref{heatpulse}). Because the feedback
is controlled through software, excellent stability in $T_0$ was
achieved by changing the feedback parameters depending on the
pulse height and temperature range.

\begin{figure}
\begin{center}
\leavevmode \includegraphics[width=.9\linewidth]{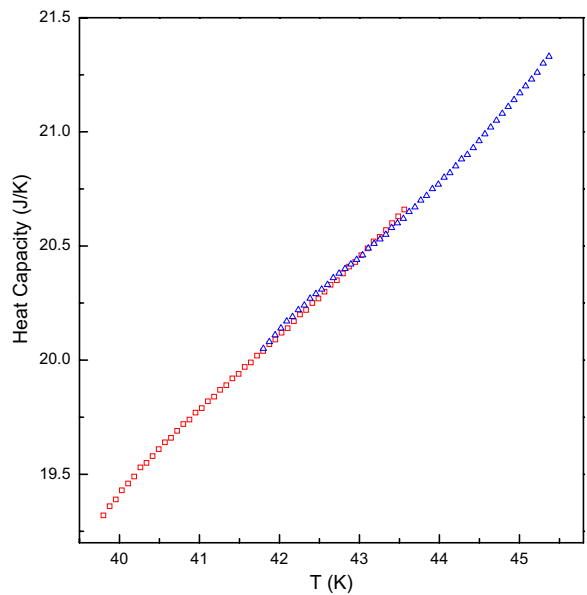}
\end{center}
\caption{Heat capacity of the calorimeter filled with pure N$_2$
sample. Data are obtained for two pulse sequences with respective
$T_0$ values separated by 2 K.} \label{adjcurves}
\end{figure}

\begin{figure}
\begin{center}
\leavevmode \includegraphics[width=.95\linewidth]{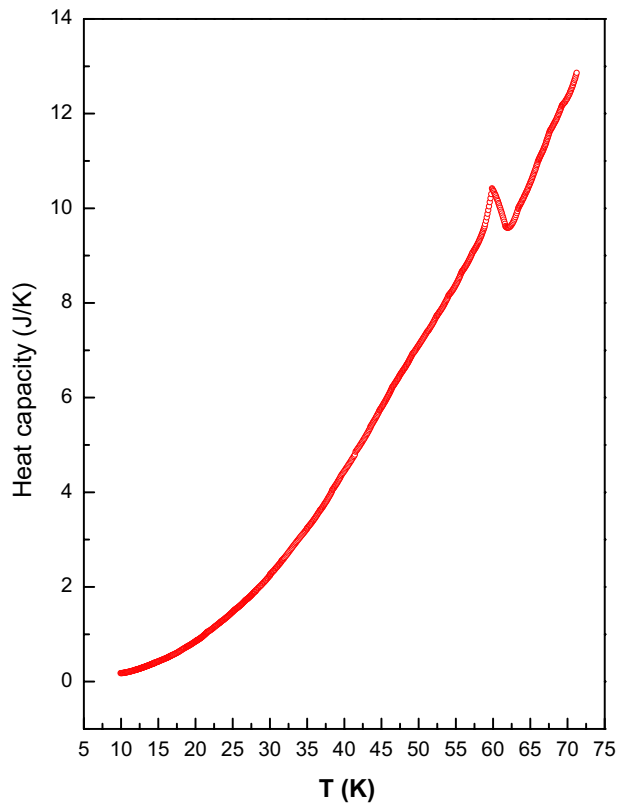}
\end{center}
\caption{Heat capacity of the empty cell with small amounts of He
exchange gas for better thermal contact. No curve fitting
techniques are utilized as the adjacent heat pulse data overlap
very well. The sharp kink at 60 K, whose origin is not known,
clearly demonstrates the high resolution of the technique near
continuous phase transitions.} \label{EcelHC}
\end{figure}

Figure \ref{adjcurves} shows the heat capacity curves of the
calorimeter filled with pure N$_2$ for two $T_0$ values.
Excellent overlap of the curves demonstrates that with this
technique, we can obtain heat capacity of samples with poor
thermal conductivity very rapidly ($\sim$3.5 K range in 3 h.),
without further curve fitting generally required with other
techniques.\cite{Xu} By extending the LabVIEW program written for
controlling the reservoir temperature $T_0$, we fully automated
the data acquisition process to obtain pulse sequence data for
various $T_0$ values with an interval of 2.5 K to ensure data
overlap from adjacent pulse sequences. Once the data are
obtained, standard third degree polynomial fits (only for the
useful temperature interval similar to the one between dashed
lines in Fig.\ \ref{heatpulse}) are employed for obtaining time
derivatives in Eq.\ (8). Once the heat capacity is calculated for
each pulse sequence, no further curve fitting is necessary and in
general for $T<30$ K, the overlap between adjacent curves is
better than the one shown in Fig.\ \ref{adjcurves}. Figure
\ref{EcelHC} shows the heat capacity of the empty calorimeter
(with small amounts of He gas for better thermal contact)
obtained through the above automated process in the $10-70$ K
range. The kink at 60 K, the presence of which is accidental,
shows the heat capacity near a continuous phase transition with
excellent resolution. The data shown without further curve
fitting clearly demonstrates the high sensitivity of the
technique.

In summary, we have demonstrated the use of a novel technique to
study the heat capacity of moderately large samples with poor
thermal conductivity in the 7.5-70 K range. A fully automated
calorimeter for rapid measurement of the heat capacity of
condensable gases utilizing the above technique has been
presented. The technique along with the automated calorimeter
with a provision to apply external electric and magnetic fields
is particularly useful for the study of continuous phase
transitions in molecular solids as well as field induced changes
in the heat capacity.

\acknowledgments The authors gratefully acknowledge the support
of L.\ Phelps, G.\ Labbe, B.\ Lothrop, W.\ Malphurs, M.\ Link,
E.\ Storch, T.\ Melton, S.\ Griffin, and R.\ Fowler. This work is
supported by a grant from the National Science Foundation No.\
DMR-962356.

\end{document}